\documentclass{article}
\usepackage{latexsym}
\usepackage{graphicx}
\usepackage{mathptmx}
\usepackage[T1]{fontenc}
\usepackage[frozencache,cachedir=.]{minted}
\usepackage[a4paper, total={6in, 8.2in}]{geometry}

\usepackage{amsmath}
\usepackage{amsfonts}
\usepackage{amssymb}
\usepackage{amsbsy}
\usepackage{amsthm}
\usepackage{authblk}

\usepackage[pdftex,colorlinks=true,urlcolor=blue,citecolor=black,anchorcolor=black,linkcolor=black]{hyperref}

\newtheoremstyle{wsc}
{3pt}
{3pt}
{}
{}
{\bf}
{}
{.5em}
{}

\theoremstyle{wsc}

\begin{document}

%
%

\title{Coarse-grained simulations of DNA and RNA systems with oxDNA and oxRNA models: Introductory tutorial}
\date{}
\author[1]{Michael Matthies}
\author[1,2]{ Matthew L. Sample}
\author[1]{Petr \v{S}ulc}
\affil[1]{School of Molecular Sciences, Arizona State University, Tempe, AZ 85281, USA}
\affil[2]{Fulton School for Engineering of Matter, Transport, and Energy, Arizona State University, Tempe, AZ 85281, USA}


\maketitle

\section*{ABSTRACT}

We present a tutorial on setting-up the oxDNA coarse-grained model for simulations of DNA and RNA nanotechnology. The model is a popular tool used both by theorists and experimentalists to simulate nucleic acid systems both in biology and nanotechnology settings. The tutorial is aimed at new users asking "Where should I start if I want to use oxDNA". We assume no prior background in using the model.  This tutorial shows basic examples that can get a novice user started with the model, and points the prospective user towards additional reading and online resources depending on which aspect of the model they are interested in pursuing.


\section{INTRODUCTION}
\label{sec:intro}
The fields of DNA and RNA nanotechnology use designed nucleic acid strands to self-assemble nanoscale structures and devices. In the past forty years since the ground-breaking work by Ned Seeman that has established the field \cite{seeman1982nucleic}, structures of increasing complexity and sizes have been experimentally realized. The promising applications of the field include diagnostics, therapeutics, molecular computation and templating for nanoscale assembly \cite{seeman2017dna,pinheiro2011challenges}. One of the most commonly used constructs in the DNA nanotechnology field is DNA origami \cite{rothemund2006folding}. DNA origami consists of a long (7000 bases) single-stranded DNA scaffold strand, and shorter (order of tens of nucleotides) staple strands that connect different domains together. Computer simulations can provide important information into the function of these devices, but also presents multiple challenges. The substantial system sizes ( encompassing up to tens of thousands of nucleotides) as well as the occurrence of rare events that may be crucial to the function of the device (e.g. breaking and creation of base pairs) present significant modeling challenges. Hence, coarse-grained models, where each particle represents multiple atoms \cite{sengar2021primer,maffeo2020mrdna}, are often employed as opposed to models with  atomistic-resolution. 

Among the leading coarse-grained models for DNA and RNA nanotechnology are oxDNA and oxRNA \cite{ouldridge2011structural,vsulc2012sequence,vsulc2014nucleotide,snodin2015introducing}, which have been parametrized to reproduce basic structural, thermodynamic and mechanical properties of DNA and RNA. Where available, the models have been shown to be in good agreement with available experimental data. They  have been utilized for nanostructure design and verification, as well as for studies of active DNA devices such as walkers, as well as for biophysical properties of DNA and RNA \cite{doye2013coarse,sengar2021primer}. In this tutorial, we describe the basics of the models, and offer a walk-through for setting up a simulation. It is meant for users who are new to nucleic acid nanotechnology modeling and are wondering what are the good resources to get started. It is beyond the scope of the tutorial to cover all aspects of molecular simulation and modeling for DNA nanotechnology and all the capabilities of the oxDNA models and associated tools. For a more exhaustive understanding of the oxDNA modeling tools ecosystem, we direct readers to additional references. 

\section{The oxDNA model and when to use it}
\label{sec:model}
\begin{figure}[htb]
{
\centering
\includegraphics[width=0.50\textwidth]{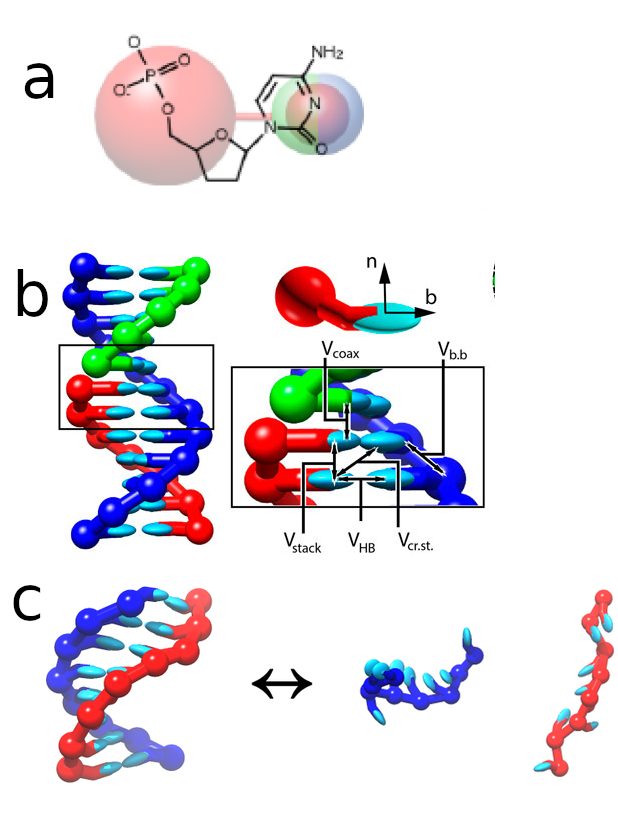}
\caption{A schematic overview of oxDNA and oxRNA models. a) A superposition of coarse-grained and atomistic nucleotide representations. Even though for visualization purposes we draw the hydrogen bonding and stacking base site as an ellipsoid, the model represents them as spherical excluded volume sites. b) A schematic of a DNA duplex, along with a representation of a single base with position vectors $\mathbf{b}$ and $\mathbf{n}$, as well as a schematic illustration of the interactions in the model (Debye-Huckel repulsion between backbone sites is not shown). c) A schematic of a melting transition in the oxRNA model. \label{fig:model}}
}
\end{figure}

OxDNA/oxRNA are coarse-grained models primarily used for simulations of DNA and RNA nanotechnology. The models are highly coarse-grained, with 1 rigid body representing the entire nucleotide (see Fig.~\ref{fig:model}). The beads have multiple interaction sites, and the interaction between the sites are empirically parameterized to reproduce DNA and RNA hairpin and duplex melting as given by the SantaLucia and Turner nearest-neighbor models \cite{santalucia1998unified,mathews1999expanded}. The models also capture the transition from single strands to duplex or stem formation, and represent the mechanical properties of highly flexible single-strands, while duplex regions behave as extensible worm-like chains, in good agreement with known mechanical properties of DNA and RNA \cite{ouldridge2011structural,vsulc2014nucleotide,snodin2015introducing}. The schematic illustration of the oxDNA model and its interactions is shown in Fig.~\ref{fig:model}. More recently, we also introduced the ANM-oxDNA model, which represents proteins alongside the oxDNA or oxRNA model. The description of the ANM-oxDNA model is beyond the scope of this tutorial, and instructions on setting-up such simulations can be found in \cite{bohlin2022design,procyk2021coarse}.

While detailed description of the model's parametrization and their properties of provided elsewhere \cite{ouldridge2011structural,snodin2015introducing,vsulc2014nucleotide,vsulc2012sequence}, the novice user should be aware of the following:  1) The oxDNA model comes in two versions, oxDNA1 and oxDNA2. The latter, oxDNA2, is the  recommended choice as it  includes salt effects using the Debye-Huckel potential and accounts for major and minor grooving, which oxDNA1 did not capture. Similarly, the oxRNA2 also includes the Debye-Huckel potential as opposed to oxRNA1, which did not support it, and hence is also the recommended interaction to use. 2) Both models can be used with two options: sequence-independent and sequence-dependent. The sequence independent version is the default version. While only A-T (A-U) and C-G base pairs can form in the sequence-independent model, their strength is the same: The model uses averaged value for all of them. This model is suitable for studies where one does not want the results to be occluded due to a particular choice of interaction strength, or where the sequence effects average out due to similar frequency of use of AT and GC base pairs. The sequence dependent model uses stacking and base pairing interaction strengths parameterized to the melting temperature predictions of duplexes. Additionally, in the oxRNA model, it also allows formation of wobble (G-U) base pairs.  3) OxDNA is the name of the coarse-grained DNA model, but is also used interchangeably to refer to the oxDNA simulation package \cite{poppleton2023oxdna}, which is a molecular simulation software package that currently implements over 20 different coarse-grained models, including oxDNA and oxRNA. Originally, the simulation package only supported the oxDNA model, hence the names were synonymous. However, since the package was first introduced as an open-source software in 2012, new models have been added to it. Additionally, the oxDNA and oxRNA models have also been implemented in the general-purpose LAMMPS simulation package \cite{henrich2018coarse}. In this tutorial, we only focus on the simulation of oxDNA with the oxDNA simulation package, which also includes additional set of scripts in Python for data evaluation.

Prior to starting any simulations,it is important to consider if the oxDNA tool can provide valuable insight into the problem you are studying. Namely, one should ask if the same research problem can be answered with a simpler and faster method, and if the process one is interested in is accurately captured by the model, or would answering it require a more detailed / accurate description of the system? Often, one can use secondary structure prediction tools such as ViennaRNA or NUPACK \cite{zadeh2011nupack,lorenz2011viennarna} for fast enumeration of possible interaction states. As opposed to these tools, oxDNA provides a three-dimensional representation, can be used to study kinetics, and simulation of complex 3D structures including pseudoknots. On the other hand, the model lacks atomistic-level resolution and cannot represent interactions with other biomolecules, which limits its applicability to studies of complex biological environments.

Currently, OxDNA and oxRNA are the only models for DNA and RNA nanotechnology that can correctly and efficiently capture single-stranded to double-stranded transitions. This makes them especially useful to study systems where base-pair creation or breaking can occur, as well as for structures which consist of both single-stranded and double-stranded regions. The model has been shown to be able to study systems consisting of up to 1 million nucleotides \cite{rovigatti2022simple}. However, simulations of diffusion-limited processes that rely on the diffusion of  single-stranded DNA or RNA, can be quite time-consuming. Currently, the model is not fast enough to e.g. simulate assembly of DNA origami structure from individual strands, and advanced biasing methods and significant computational resources need to be applied to study systems of such sizes \cite{snodin2016direct}.

The simulation package implements both CPU and GPU molecular dynamics engines, and CPU implementations of the Monte Carlo and cluster Virtual Move Monte Carlo (VMMC) algorithms. The latest version of the code is freely available at \url{https://github.com/lorenzo-rovigatti/oxDNA}. The online repository features code documentation, which includes installation instructions and also provides a more comprehensive explanation of the model and its parameters. In the following, we walk through the steps of setting up and evaluating an example simulation. It is beyond the scope of this tutorial to cover all aspects of molecular simulations, and for the general techniques and explanation of theory behind molecular simulation, we recommend popular textbooks such as \cite{frenkel2001understanding,tuckerman2010statistical}. A more condensed introduction to molecular simulation algorithms is also provided in \cite{munsky2018quantitative}.

\section{Setting-up and running simulations}

\subsection{Creating the starting configuration}
The first step in setting-up a simulation is the preparation of a system in the oxDNA model file format. The oxDNA (or oxRNA) models requires two files: the topology file (typically ending with .top suffix) and configuration file (typically saved as a .dat file). Below, we list an example of the topology file for a system consisting of two DNA strands, each of length 3 bases:
\begin{verbatim}
6 2
1 T -1 1
1 A 0 2
1 C 1 -1
2 G -1 4
2 T 3 5
2 A 4 -1
\end{verbatim}
where the first line indicates there are 6 nucleotides in two different strands. Then, the sequence is listed in the $3'$ to $5'$ order, with one nucleotide per line. The first column is the id of the strand the nucleotide belongs to, then the base type, and the id of the $3'$ and $5'$ neighbor. Alternatively, it is possible to replace the base type (A,T,C,G) by a number with absolute value greater than 10, in which case it can only bind to a base with id such that their sum is 3. For example, in the following topology the first nucleotide can be only able to bind to the last one, and no other base pairs are possible for it, thus avoiding some possible undesired mismatches that you want to prevent from forming:

\begin{verbatim}
6 2
1 13 -1 1
1 A 0 2
1 C 1 -1
2 G -1 4
2 T 3 5
2 -10 4 -1
\end{verbatim}

Future stable release of the oxDNA tool will also support a new topology format where the sequence is simply listed as one strand per line, but the above format will remain supported for backward compatibility. An example of the first three lines of the configuration file are 
\begin{verbatim}
t = 0
b = 20.0 20.0 20.0
E = -1.27  -1.42 0.14
\end{verbatim}
where the first line is the timestep of the simulation (starting at 0), the second line is the dimensions of the simulation box edges, and the third line is the total, potential, and kinetic energy per particle respectively. It is advised that the box size is large enough such that the simulated structure cannot interact with itself (e.g. has to be larger than the largest distance between any pairs of nucleotides in the system). This can be easily verified by visualizing the structure in oxView tool \cite{bohlin2022design} and switching on the Box option in View tab. Then, after the top three header lines, the configuration file has one line per particle (not shown above). Each line lists, for each nucleotide in separate columns, x, y and z position, the three components of the particle vectors $\mathbf{b}$ (pointing from the center of mass of nucleotide to base site) and $\mathbf{n}$ (perpendicular to $\mathbf{b}$ and will point towards $5'$ neighbor in a formed duplex, see Fig.~\ref{fig:model}b). Then, the three components of the velocity and angular velocity vector are listed as well.
Once the simulation is run, the simulation code will also produce a trajectory file, which is a series of configuration files (in the same format as described above) concatenated one after another.

It would be too tedious to manipulate and create the topology and configuration files directly, so a set of tools are provided to convert from common design tools such as CADNano, Tiamat or PDB format \cite{douglas2009rapid,williams2009tiamat}. Some other existing tools, such as MagicDNA or sCADNAno \cite{huang2021integrated,doty2020scadnano}, also allow directly saving designed system in oxDNA format. 

The standard tool for oxDNA visualization, oxView, is available as a browser-based application at \url{https://oxView.org}. It can import from common DNA nanotechnology formats directly into oxDNA, and then save the system as .top and .dat files in the format above. Importing structures from other design tools is advised for origami sized systems, while smaller systems (like tiles) can be created directly using the oxView editing interface, which allows creation and deletions of single-stranded and double-stranded sections and connecting them together. The oxView tool along with its use for structure import and generation is described in detail in \cite{bohlin2022design}. The conversion tool between different formats is also provided as a standalone website \url{http://tacoxdna.sissa.it/} \cite{suma2019tacoxdna}.

\subsection{Installation of the oxDNA tool}
The software tool is intended to be run in a high performance computing (HPC) environment, such as a dedicated workstation or computer cluster. Such systems typically run using Linux systems, and as such, our tool is currently intended to be run in a Linux environment. It can be also run in a Linux virtual system on Windows. To compile the executable (binary) of the program, go to the oxDNA directory after downloading the tool and type
\begin{verbatim}
    mkdir build
    cd build
    cmake ../ -DCUDA=1
    make 
\end{verbatim}
If successful, the above command will lead to a binary file called \texttt{oxDNA} in the build directory. The above commands also assumes you have the CUDA toolkit library installed on your system that supports GPU computing. If you leave out the \texttt{-DCUDA=1} parameter, the code will only be compiled with support for a single CPU simulations. In most HPC settings, one needs to import a NVIDIA CUDA compiler and a compatible C++ compiler (we recommend GNU C Compiler) using the \texttt{module add} command. Please refer to your computer cluster documentation for details how to import the right libraries.

Optionally, you can install Python binding libraries as well, which gives you access to the oxpy modules for preparing and running the simulations from a Python interface, as well as providing access to the oxDNA Analysis Tools modules for analysis of simulations in Python.  In that case, the installation process is
\begin{verbatim}
    mkdir build
    cd build
    cmake ../ -DCUDA=1 -DPython=1 -DOxpySystemInstall=1
    make 
    make install
\end{verbatim}
The above Python binding will only work with Python version 3.9 or higher. For future reference of troubleshooting of known installation issues, please refer to the online documentation of the tool. For additional information and troubleshooting, follow the instructions on the github repository documentation page \url{https://lorenzo-rovigatti.github.io/oxDNA/}. 

\subsection{Use of online resources to run oxDNA}
 As an alternative to installing and compiling your own oxDNA code locally,  online tools are available to perform simulations for you.
To simplify the setup and use of the model, we maintain a free online webserver \url{https://oxDNA.org}, where the topology and configuration files can be directly uploaded. The server is equipped with 8 GPU cards and supports oxDNA, oxRNA and ANM-oxDNA simulations. It is meant for users that do not have access to their own computer clusters and are primarily interested in running unbiased simulations of DNA or RNA systems. The details of how to use this service is provided in \cite{poppleton2021oxdna}. For certain applications this might be a better option, as the server interface is optimized for typical use cases of the oxDNA and oxRNA models. It provides basic online analysis tools that can produce mean structures, preform flexibility analysis, create videos of simulated trajectories, show bond occupancy, and calculate distributions of angles between different helices and distances between groups of nucleotides. In the rest of the tutorial, we assume that you are running all simulations on a Linux cluster, and hence need to setup all simulations by yourself.

\subsection{Setting up simulation parameters}

\subsubsection{Molecular Dynamics on CPU}
Once you have prepared the starting configuration and topology, we need to setup a text file with instructions to run the simulation. This file is typically named \texttt{input}, and contains parameters of the simulations. An example of such a file is provided below:

\begin{verbatim}
####  PROGRAM PARAMETERS  ####
backend = CPU
seed = 4982

####    MD PARAMETERS    ####
sim_type = MD
newtonian_steps = 103
diff_coeff = 2.50
thermostat = john
dt = 0.003
verlet_skin = 0.05

##### MODEL PARAMETERS #####
interaction_type = DNA2 
use_average_seq = no
seq_dep_file = oxDNA2_sequence_dependent_parameters.txt
salt_concentration = 1.0
T = 25C

####    INPUT / OUTPUT    ####
steps = 1e8
topology = 75deg.top
conf_file = 75deg.dat
trajectory_file = trajectory.dat
refresh_vel = 1
time_scale = linear
restart_step_counter = 1
energy_file = energy.dat
print_conf_interval = 1e5
print_energy_every = 1e4
\end{verbatim}

Any line that starts with {\tt\#} are treated as a comment and are ignored by the program. The simulation parameters are grouped by their role. First, \texttt{backend} specifies we will run the code on CPU, and \texttt{seed} initializes the random number generator. Omitting the \texttt{seed} parameter prompts the program to default to a randomly generated number when launched. If you run one structure in two different simulations with the same seed, they will produce identical trajectories, so make sure to launch simulations with unique seeds. Next, we set the parameters of the molecular dynamics (MD) algorithm. While detailed description of molecular dynamics is available elsewhere \cite{frenkel2001understanding}, it can be approximately understood as numerical integration of Newton's equations of motion. This involves parameters like timestep \texttt{dt} and verlet list, which is important for neighbor list calculation, as elaborated in \cite{frenkel2001understanding}). Each nucleotides movement in the simulation is dictated by forces equal to the derivatives of the oxDNA model potential. Additionally, it is coupled with a thermostat, which ensures that the states sampled by the system correspond to the specified temperature. In the example above, we use an Andersen-like thermostat (specified by parameter \texttt{john} in reference to the article which introduced it \cite{russo2009reversible}). This thermostat periodically resamples random velocities and angular momenta of nucleotides so that they correspond to the Boltzmann distribution at the specified temperature.  The parameters of the thermostat are \texttt{diff\_coeff} which sets the diffusion constant of a single nucleotide, and \texttt{newtonian\_steps} which is related to the frequency with which the velocities of nucleotides are refreshed. It is advised that these parameters are kept to their values specified above. The \texttt{steps} parameters specifies how long the simulation should run (i.e. how many steps it will perform). This number is system-dependent, and it should be as large as necessary to sample all relevant states of the simulated system. 

The model parameters section specifies that we use the oxDNA2 model version (\texttt{interaction\_type = DNA2}), with its sequence-dependent variant. The file \texttt{oxDNA2\_sequence\_dependent\_parameters.txt} must either be copied from the main oxDNA directory to the directory where the simulation is launched or the full path to the file location can be provided. The above configuration sets the sodium concentration to 1M using the Debye-H\"{u}ckel potential, and the temperature to 25 Celsius degrees. An identical set of parameters would also apply to the oxRNA model, however the interaction is called RNA2. Additionally, there is an optional (but highly recommended) parameter \texttt{mismatch\_repulsion = 1}, which introduces a weak repulsion between mismatched base pairs to correct for the high stability of mismtached base pairs in the oxRNA model \cite{vsulc2014nucleotide}.

Finally, the input/output section specifies the name of the topology and starting configuration files (\texttt{75deg.top} and \texttt{75deg.dat} in this example). The generated trajectory will be saved to \texttt{trajectory.dat} (specified to be saved every $10^5$ simulation steps). The energy per particle as a function of time will be saved to a file named \texttt{energy.dat} every $10^4$ steps, using linear time scale. Setting \texttt{restart\_step\_counter} to 1 means that every time the simulation is stated, it deletes the \texttt{energy.dat} and \texttt{trajectory.dat} files and then restarts the simulation from time 0. If the parameter is set to 0, the simulation will instead continue appending to existing trajectory.dat and energy.dat files, resuming from the time specified in the configuration file.  The simulation saves the last configuration when it ends (or the job is cancelled or killed by the server) to a file named \texttt{last\_conf.dat}. If you use the option \texttt{restart\_step\_counter = 0}, it is vital to ensure you are restarting the simulation from the last simulation state by setting \texttt{conf\_file = last\_conf.dat}. 

To run the simulation, it is advised to create a separate directory which contains the topology, configuration and input files. The simulation can then be launched by 
\begin{verbatim}
$OXDNAPATH/build/bin/oxDNA input
\end{verbatim}
where \$OXDNAPATH is set to the oxDNA installation directory. The program will then run and produce the output files in the same directory. To ensure you simulation is running as intended, it is advised to observe the values of energy printed on the screen (ideally they should be about $-1.4$ per nucleotide for a system consisting of mostly base-paired nucleotides), and to use oxView to visualize the produced trajectory.

\subsubsection{Molecular Dynamics on GPU}
Single CPU simulations are usually too slow for most systems of interest. If the size of the studied system exceeds about 300 nucleotides, the GPU simulation can provide up to two orders of magnitude speed-up. To use a GPU-version of oxDNA, the first two sections of the input file have to be changed to

\begin{verbatim}
####  PROGRAM PARAMETERS  ####
backend = CUDA
backend_precision = mixed
use_edge = 1
edge_n_forces = 1

####    MD PARAMETERS    ####
sim_type = MD
CUDA_list = verlet
verlet_skin = 0.2
max_density_multiplier = 10
CUDA_sort_every = 0
newtonian_steps = 103
diff_coeff = 2.50
thermostat = john
dt = 0.003
\end{verbatim}
The detailed explanation and fine-tuning of the parameters above are given in the oxDNA code documentation. For a majority of large systems (e.g. DNA origami), these listed values should provide optimal speed on the latest GPU cards. The remaining sections setting-up the model and input and output files are the same as they were for the CPU implementation.

\subsubsection{Monte Carlo Simulation on CPU}
The simulation code also supports both the Metropolis Monte Carlo Algorithm and Virtual Move Monte Carlo (VMMC) algorithm for the oxDNA and oxRNA models. The Monte Carlo algorithm relies on proposing a trial move, either with a single particle or a group of particles, and subsequently accepting or rejecting it \cite{frenkel2001understanding}. The Virtual Move Monte Carlo algorithm \cite{whitelam2007avoiding} is particularly efficient at sampling the states of smaller systems (ideally less than 100 nucleotides), which it achieves faster than molecular dynamics. The input file for the MC algorithm (which replaces the MD PARAMETERS section of the CPU version of the code) is:
\begin{verbatim}
####    MC PARAMETERS    ####
sim_type = MC
ensemble = NVT
delta_translation = 0.10
delta_rotation = 0.25
list_type = cells
\end{verbatim}

The parameters \texttt{delta\_translation} and \texttt{delta\_rotation} set the mean size of proposed translation or rotation moves, and should ideally be chosen so that about 30\% of suggested moves are accepted. Large scale moves will be rejected too often, while short moves will be accepted, but overall it will take too many proposed moves before the system significantly changes. Provided \texttt{list\_type = cells} is one of the possible means (along with verlet lists) that algorithm can efficiently keep track of its neighbors. For the VMMC simulation, the recommended input parameters are

\begin{verbatim}
####    VMMC PARAMETERS    ####
sim_type = VMMC
ensemble = NVT
delta_translation = 0.10
delta_rotation = 0.25
verlet_skin = 1.0
small_system = 1
maxclust = 30
\end{verbatim}

In the above, there are two optional parameters: \texttt{maxclust} and \texttt{small\_system}, which in certain scenarios (typically for system sizes below 50 nucleotides) can increase the speed of the algorithm. When the simulation tool stops, it prints out benchmarking information (such as CPU time per one simulation step) which can be used to fine-tune parameters for maximum efficacy.

\subsection{Relaxation of a structure}
When using designs exported directly from design tools, it is often necessary to perform a "relaxation" simulation prior to the production run (the simulation intended to collect data about the system). This relaxation phase is crucial because design tools typically place nucleotides into positions that are not physically possible. For example, adjacent nucleotides on the same strand might be placed too far from each other, exceeding the permissible distance that the covalent bond between nucleotide backbones would allow. The oxDNA model force-field would in such a case result in a nearly infinite force trying to pull these "stretched" nucleotides together, causing numerical instability and program error. Another common issue with imported designs is that sometimes nucleotides are placed too close to each other, resulting in them overlapping. Since the repulsion forces in the oxDNA model will be enormous if two nucleotides are too close to each other, this can result in very high forces and subsequent numerical instability and program crash. It is hence recommended to perform relaxation simulations prior to production runs. We will list the parameters needed to run a relaxation simulation below (they need to be added to either CUDA or CPU simulation input file):

\begin{verbatim}
####  RELAX PARAMETERS  ####
dt = 0.001
max_backbone_force = 10
\end{verbatim}

Where \texttt{dt} is set to a smaller value as compared to the production run to increase numerical stability and \texttt{max\_backbone\_force} is the parameter permitting overstretched backbone bonds.  The relaxation steps can be run for a certain number of steps (typically $10^5$ to $10^7$, but can be longer for larger systems with many stretched regions). In some cases (which feature lots of overlapping nucleotides), it might also be necessary to first run Monte Carlo simulation before proceeding to molecular dynamics relaxation. The automated relaxation is also implemented in the oxDNA.org webserver.

\subsection{Running simulations using the Jupyter notebook interface}

Beyond the command line interface, oxDNA also supports python bindings. This capability facilitates both oxDNA simulation and analysis directly from a Jupyter notebook. The setup is described in more detail in the online documentation at \url{https://lorenzo-rovigatti.github.io/oxDNA/oxpy/index.html#an-example-of-a-simple-simulation}.  To streamline the setup of routine simulations, we provide basic boiler plate code.
The Jupyter notebook code provided below preforms the setup of a default production run simulation and provides a convenient way to view and analyse the trajectory. 

\begin{minted}{python}
# visualisation library
from oxDNA_analysis_tools.UTILS.oxview import from_path
# boilerplate code to setup and monitor simulations
from oxDNA_analysis_tools.UTILS.boilerplate import setup_simulation,\
                                                    get_default_input,\
                                                    Simulation



# get default settings for a CUDA simulation
parameters = get_default_input()
# setup sequence dependence 
parameters["use_average_seq"] = "no"
parameters["seq_dep_file"] = "../../oxDNA2_sequence_dependent_parameters.txt"

# setup simulation
# we will simulate the 75 degree layered tile from 
# https://doi.org/10.1021/jacs.8b07180
input_file = setup_simulation("./75deg.top", # topology 
                              "./75deg.dat", # starting configuraiton
                              "./simulation",         # output path
                              parameters,   # simulation parameters
                              kill_out_dir  = True)

# create a simulation class
s = Simulation(input_file)

# view initial configuration
s.view_init()

# run the simulation
s.run()
\end{minted}

This code will setup and run a production run of the 75 degree layered crossover tile. To monitor the progress of the simulation you can utilize the following code. 

\begin{minted}{python}
# visualize simulation progress
s.plot_energy()
\end{minted}

This will produce a plot of the total energy per simulation step. The red line indicates the completion of the simulation. This plotting function can be rerun as many times as needed to follow the simulation progress.
To visualize progress one can additionally utilize following code, visualizing the last configuration: 

\begin{minted}{python}
# displays the last configuration
s.view_last()
\end{minted}

Finally after the simulation finishes one can analyze the resulting trajectory using the following jupyter code, calling the oxDNA analysis tools mean script.

\begin{minted}{python}
# after the simulation is finished, 
# lets perform a mean structure analysis using oxDNA_analysis_tools
oat mean ./simulation/trajectory.dat -o ./simulation/mean.dat -d\
                                            ./simulation/devs.json
\end{minted}

The resulting files can be visualized in the Jupyter notebook using:
\begin{minted}{python}
from_path("./simulation/75deg.top",
              "./simulation/mean.dat",
              "./simulation/devs.json")
\end{minted}

We provide this code and the configuration files as a Jupyter notebook in the example folder "BOILERPLATE\_Jupyter", shipped with the oxDNA code.


\section{Advanced simulations}
The examples above covered the basic setup for unbiased simulations on GPU and CPU. For many systems, this might not be enough: If, even after few days of simulations, you have still not seen the simulated system sample the relevant states often enough, this suggests that the unbiased simulation is not providing sufficient data to be representative of the full behavior of the structure. For example, from experimental FRET measurements, you know that the arms of a DNA nanostructure sample certain ranges of possible distances. However, if the oxDNA simulations do not show the entire range of the motion, it means that it has not been run long enough, or that more advanced methods to enhance the sampling need to be employed. The oxDNA simulation implements discrete umbrella sampling in combination with the VMMC algorithm, which is an ideal tool to sample the entire free-energy landscape as a function of numbers of base pairs in the system. Examples of use include simulations of melting of duplexes, hairpins and pseudoknots and strand displacement system \cite{srinivas2013biophysics,hong2020understanding}. This method is documented, along with provided example codes, on the github repository of the oxDNA code. We also support umbrella sampling as a function of continuous parameter (typically distance between groups of nucleotides) that can be used to enhance the sampling of large scale conformational change of different structures. The example code to setup such simulations is available as part of \cite{sample2023hairygami} and is also discussed briefly below. Finally, the model also support enhanced kinetics sampling to extract e.g. rates of a particular reaction such as strand displacement. The forward flux sampling method is supported both on CPU and GPU in the oxDNA code, with examples provided in the the oxDNA documentation on github. 

\subsection{Continuous parameter umbrella sampling}
An open-source Python wrapper of the oxDNA Python bindings, available at \url{https://https://github.com/mlsample/ipy_oxDNA}, provides a semi-automated Jupyter notebook interface to run continuous parameter umbrella sampling. This runs on the GPU implementation of the oxDNA molecular dynamics engine, and as such allows for enhanced sampling of origami-sized systems or larger. The Jupyter notebook interface enables a greater ease of use and supports reproducible and open source simulation notebooks. Furthermore, NVIDIA's multiprocessing service can be harnessed to increase simulation throughput over 2.5x by allowing over 40 origami sized simulations to be run on a single GPU.


Umbrella Sampling works by adding an external biasing potential to the system to enforce sampling all values of the chosen order parameter. An order parameter is any measurable value that is of interest in a simulation. An example of a chosen order parameter can be the distance between two groups of nucleotides on the ends of a DNA origami that we expect to undergo large scale conformational change over a range of distances in the simulation. 
The biasing potential effectively 'flattens' the energy landscape, making it easier for the system to overcome energy barriers and explore a wider range of states. To ensure we span the entire range of your order parameter of interest, we run multiple simulations (windows), where each window samples only a specific subset of the order parameter.

The set of biased simulations obtained by running different windows are then combined using the Weighted Histogram Analysis Method (WHAM)  \cite{grossfield_wham}. WHAM effectively 'removes' the bias imposed by the external biasing potential applied during the simulations, and then stitches the distributions of each window together to give a single unbiased estimate of the free energy profile for your order parameter. Careful selection of the order parameter is crucial, as it should be able to effectively distinguish between the different states or conformations of the system. 

Following the import of the necessary python modules:
\begin{minted}{python}
# Import necessary modules
from umbrella_sampling import ComUmbrellaSampling
from oxdna_simulation import SimulationManager
\end{minted}
the first step to use the python wrapper is to specify the directory that contains the oxDNA conformation and topology files (file\_dir), which is ideally prepared in the command line prior to running the simulation. Next, you need to specify a unique name for the directory which will contain our umbrella sampling system (system). We can then define our order parameter, where in this example we use the distance between the center-of-mass between nucleotides in group one (com) and group two (ref), defined by their nucleotide index.
\begin{minted}{python}
# Define the directory where your files are stored
file_dir = 'abs/path/to/files'

# Define the name for your system
system   = 'user_defined_system_name'

# Define the particles for center-of-mass and reference
com_list = '1,2,3' # This is a comma-separated list of particle indices
ref_list = '4,5,6' # These indices can be found using oxView
\end{minted}
 We next set the umbrella simulation parameters which are determined \textit{a priori} or refined over umbrella simulation trials. These include the order parameter sampling range (xmin, xmax), the number of simulation windows (n\_windows), and the stiffness of the external biasing potential (stiff).
\begin{minted}{python}
# Define the parameters for the umbrella sampling
xmin = 0         #Lowest com distance attemped sampling
xmax = 10        #Highest com distance attemped sampling
n_windows = 20 #Parallelizable simulations sampling subsets of range
stiff = 1        #Stiffness or 'k' value of umbrella potential
\end{minted}
For the oxDNA simulation parameters, we need to set the number of steps for each window, the temperature, and the intervals at which to print the energy and configuration data. We can also modify other oxDNA parameters as required, such as salt concentration.
\begin{minted}{python}
# Define the parameters for the equilibration phase
equilibration_parameters = {'steps':'1e7', 'T':'20C',
                            'print_energy_every': '1e6', 'print_conf_interval':'2e6'}
# Define the parameters for the production phase
production_parameters = {'steps':'2e7', 'T':'20C',
                             'print_energy_every': '2e6','print_conf_interval':'1e7'}

# Define a list of shared parameters
shared_params = [simulation_manager,  n_windows, com_list,
                    ref_list, stiff, xmin, xmax]  #optional list to reduce reduencay
\end{minted}
We create an umbrella sampling object and a simulation manager object. If a system directory doesn't exist, instantiating a ComUmbrellaSampling object will create one. If a system directory already exists, it will read the current state of the directory, and will "reattach" to the directory. This means if we need to clear the Jupyter kernel or start a new session with a umbrella simulation run previously, we can  "reattach" to the umbrella system directory and preform analysis or continue simulations without modifying the contents or overwriting files. The simulation manager allocates the simulation to the available GPU and CPU resources optimally within on a single node.
\begin{minted}{python}  
# Initialize the umbrella sampling and simulation manager objects
us = ComUmbrellaSampling(file_dir, system)
simulation_manager = SimulationManager()
\end{minted}
Before the simulations are built, we start the NVIDIA multiprocessing service for better performance.
\begin{minted}{python}
# Enable the Nvidia multiprocessing service
simulation_manager.start_nvidia_cuda_mps_control()
\end{minted}
Now, you'll prepare and run equilibration simulations. These will generate your simulation windows to their assigned sampling range. It is recommended to start with a pre-equilibration check, a brief 50,000 step simulation, to verify your system setup.

If the pre-check is successful, re-run the 'build\_equilibration\_runs()' method. This will overwrite your previous equilibration setup, emphasis on rebuilding will delete previous files, and start the true equilibration process. The number of steps for this phase depends on your system and order parameter, but for origami-sized systems, a minimum of 2 million steps is suggested.

The 'simulation\_manager.run()' method executes these equilibration simulations as a set of 'grandchild' processes. This lets you continue using your notebook while simulations are running. However, if you prefer to block your kernel while running umbrella simulations (necessary for a umbrella sampling python script with Jupyter-validated umbrella setups), use the 'simulation\_manager.worker\_manager()' method instead. This spawns the simulations as a set of 'child' processes.
\begin{minted}{python}
# Build and run the equilibration simulations
us.build_equlibration_runs(*shared_params, equilibration_parameters)
simulation_manger.run() # This runs the equilibration simulations
\end{minted}
As the simulations run (or afterwards), you can analyze the time series order parameter values using the following commands:
\begin{minted}{python}
# Analyze the simulation results while they're running or afterwards
us.com_distance_observable(com_list, ref_list)
plt.figure(dpi=200)
for idx in range(0,n_windows,1):
    us.analysis.view_observable('eq', idx, sliding_window=False)
plt.legend(fontsize=4)
\end{minted}
You can view the final conformation of any window by running the following:
\begin{minted}{python}
# View the last conformation of any window
us.equilibration_sims[idx].analysis.view_last()
\end{minted}
Finally, you can proceed to run the production simulation and the Weighted Histogram Analysis Method (WHAM) analysis:
\begin{minted}{python}
# Build and run the production simulation and WHAM analysis
us.build_production_runs(*shared_params, production_parameters)
simulation_manger.run()

wham_dir = os.path.abspath('../wham/wham')
n_bins = '200'
tol = '1e-5'
n_boot = '30'
us.wham_run(wham_dir, xmin, xmax, stiff, n_bins, tol, n_boot)
\end{minted}
Finally, after you have run the WHAM technique, you can view your free energy profile by running:
\begin{minted}{python}
# View your free energy profile
us.plot_free()
\end{minted}

A plethora of Jupyter notebooks and python scripts are provided in the github repository as well.

\section{SIMULATION ANALYSIS}
After simulations have been completed, it is often necessary to extract the relevant information using statistical analysis. Each case is system-specific, but we provide towards this goal
oxDNA Analysis Tools (oat), a Python-based suite designed for analyzing oxDNA simulations. These tools cover some typical use-cases for analysis of simulations, and can also be modified by the users to suite their current needs. They can be used as a series of standalone command-line scripts or as importable Python modules for creating personalized analysis tools. The command-line scripts can be accessed by the command oat <script name> <script arguments>, with autocomplete features available for script names. The suite includes a diverse range of scripts for various functions like bond analysis, clustering, and principal component analysis, among others. Additionally, the scripting API for oat can be imported into Python scripts, providing utilities such as the optimized oxDNA trajectory reader. Comprehensive API documentation is available for both top-level analysis and utility APIs at \url{https://lorenzo-rovigatti.github.io/oxDNA/oat/index.html}.

\section{SIMULATION VISUALIZATION}
The oxDNA viewer, or oxView, is a robust, browser-based tool for visualizing and editing oxDNA configurations. Using the Three.js JavaScript library, it provides a seamless experience for handling large configuration files, with the capacity to manage up to 1.2 million nucleotides. Its interface is intuitive, allowing users to simply drag and drop a topology, configuration, or trajectory files into the browser window. It also supports JSON overlay files and offers a variety of features, including video options, different output formats, console commands, and APIs for scene and edit operations. Additionally, it enables rigid body simulations and 3D printing exports.

Notably, oxView incorporates a "relaxation" simulation, a necessary step prior to the actual data collection to correct potential physical inconsistencies in nucleotide placements, which can replace the necessity to run relaxation simulation from the command line as described above. For a more detailed understanding of oxView's capabilities and applications, see \cite{bohlin2022design}.

\section{FUTURE DEVELOPMENT}
The simulation code is still in an active development, and as such further extensions (such as inclusion of more accurate coarse-grained representation of proteins, nanoparticles, support for DNA-RNA hybrids and more accurate representation of different salt conditions, sequence-dependent mechanical properties, capture of hydrodynamic interactions in kinetic simulations) are either in active development or on the feature-request lists. Similarly, should more experimental data become available (e.g. thermodynamic measurements of melting temperatures of large DNA nanostructures, kinetic measurements of strand displacement reactions and single molecules experiments on duplex as well as single-stranded DNA or RNA), it will present opportunity for application of novel fitting methods and verification algorithms to better parameterize the model. 

At the same time, the detailed oxDNA and oxRNA simulations can be used to parameterize Markov chain models (e.g. \cite{zolaktaf2023predicting}) that describe interactions between DNA and RNA strands in a state space defined by number of base pairs between strands rather than in terms of x-y-z coordinates in 3D space as molecular models do.  Where limited experimental data are available, the oxDNA and oxRNA models can be used to augment the training dataset. 

Every new release of oxDNA aims to preserve backward compability, even if new features are added that might make some of the older approaches obsolete. To get in touch with feature request or ask for help with exisiting code, the developers can be contacted at \url{https://github.com/lorenzo-rovigatti/oxDNA/issues}.

\section*{ACKNOWLEDGMENTS}
This material is based upon work supported by the National Science Foundation under Grant DMR-2239518. We acknowledge contributions of Thomas Ouldridge, Flavio Romano, Ben Snodin, Erik Poppleton, and Lorenzo Rovigatti to the oxDNA model and simulation and analysis tools development.

\vspace{6pt}

\footnotesize

\bibliographystyle{old-mujstyl}

\bibliography{refs}


\end{document}